# Platformization of Inequality: Gender and Race in Digital Labor Platforms (Pre-Print)

*Isabel Munoz, Syracuse University*
*Pyeonghwa Kim, Syracuse University*
*Clea O'Neil, Skidmore College*
*Michael Dunn, Skidmore College*
*Steve Sawyer, Syracuse University*

We contribute empirical and conceptual insights regarding the roles of digital labor platforms in online freelancing, focusing attention to social identities such as gender, race, ethnicity, and occupation. Findings highlight how digital labor platforms reinforce and exacerbate identity-based stereotypes, bias and expectations in online freelance work. We focus on online freelancing as this form of working arrangement is becoming more prevalent. Online freelancing also relies on the market-making power of digital platforms to create an online labor market. Many see this as one likely future of work with less bias. Others worry that labor platforms' market power allows them to embed known biases into new working arrangements: a platformization of inequality. Drawing on data from 108 online freelancers, we discuss six findings: 1) female freelance work is undervalued; 2) gendered occupational expectations; 3) gendered treatment; 4) shared expectations of differential values; 5) racial stereotypes and expectations; and 6) race and ethnicity as an asset. We discuss the role of design in the platformization and visibility of social identity dimensions and the implications of the reinforced identity perceptions and marginalization in digital labor platforms.

## 1 INTRODUCTION

In 2022, 60 million workers, or 39% of the US workforce, turned to freelancing as large swaths of workers redefine what work means beyond traditional career paths [69]. Amid the ongoing growth of online freelancing in the US, we investigate how social identities are embedded within the digital platforms that enable these arrangements and illuminate the implications for the growing number of online freelancers. Beyond brokering employment transactions, digital labor platforms are reshaping employment relationships in part by controlling and commodifying online freelancers' presentations of self. Research makes clear that freelancers' digital identities on these platforms are being constrained by profile structure, client ratings, and the platform's algorithms and norms [47]. Yet, even as digital labor platforms are expected to continue their growth trajectory, we lack a thorough understanding of how gender, race, and ethnicity impact freelancers' platform experiences and strategies in online freelancing. This article contributes to conceptualizations of identity and the role of platforms and algorithms in the gig economy, which has been explored across existing CSCW and CHI scholarship [14,18,24,37,43,47].

      Here we argue that online freelancing enables the platformization of race, gender and other social identity attributes, leading to continued work inequality. Here, 'platformization' refers to the process of defining workers' various identity dimensions like gender and race through the social and technical elements in a platform. For example, freelancing platforms typically specify terms of use and a profile structure that determines how workers can perform their identity on each platform. This means freelancers have to abide by the rules and standardization of identity if they wish to find and conduct freelance work on these platforms.

We focus on Upwork[1] as the digital labor platform for the research reported on here. Online freelancing, enabled by Upwork and other platforms, is one form of gig work that is fully remote, relies on advanced skills, and requires extended interaction between workers and clients [47,61]. Three reasons make online freelancing an important domain to explore the platformization of identity attributes like gender and race: 1) the growing importance of this form of work; 2) the unique opportunities it provides for future workers, including access to more specialized and flexible work; and 3) the gap in research about the role of platforms in addressing, transforming or exacerbating marginalization and inequality in this form of work. This work is guided by the following research question: *Does platform-mediated work mediate, perpetuate or exacerbate the marginalization that workers typically experience in traditional labor markets?*

To date, the marginalization debate is framed by two contradictory perspectives. Idealist accounts of online freelancing view this form of work as an antidote to many issues raised in the traditional labor market, emphasizing its role in promoting equitable opportunities for workers who may otherwise be excluded. Positive impacts of this form of work include 1) equalizing traditional labor market for workers from historically marginalized groups by mitigating exclusion based on race, gender, and class, and disability [19,27,53]; 2) lowering the entry barrier to the labor market through reducing transaction costs in worker-client matching [22]; 3) empowering underemployed workers by allowing them to sell their skills to clients worldwide [27]; and 4) facilitating a reconciliation of work and non-work spheres through flexibilization of spatio-temporal barriers for work [8,50].

In contrast, critical views highlight that online freelancing reproduces extant issues in the traditional labor market and introduces new ones, exacerbating working conditions for the workers from historically marginalized groups, especially for women and racial minority workers [7,21,36,48]. These issues include 1) the gender pay gap [24,25,32]; 2) the voluntary or involuntary participation of women in unpaid care and domestic work [3,36]; 3) discriminatory treatment from algorithms and affordances on digital labor platforms [2,43,49,60]; 4) discriminatory treatment from clients [16,30,43,53]; and 5) occupational segregation based on gender [21,26,42,48].

To successfully adjudicate between these perspectives, we must center the experiences of the workers in online freelancing, paying close attention to the role of social identities such as gender and race in the socio-technical ecosystem of digital labor. That is, we must investigate the role of social and technical elements in the "platformization" of gender and race to understand the broader implications for online freelancers. To do this, the study reported on here builds from what is currently known about workers' differential experiences with online work, their work trajectories, and the roles of freelancing platforms in limiting or exacerbating marginalization. As detailed below, we conducted a longitudinal, panel study with 108 US-based online freelancers across three years of annual data collection. We contribute empirical and conceptual insight about the role of platforms in differential challenges and strategies among online freelancers, by focusing attention to social identities such as gender, race, ethnicity, and occupation. Findings provide specific insight into the roles that freelancing platforms can play in addressing, exacerbating or transforming known issues of inequality among workers from marginalized populations [5,30,63].

---

[1] See: https://www.upwork.com/.

## 2 LITERATURE REVIEW

### 2.1 Online Freelancing and Women and Minority Workers

Similar to other types of gig work, online freelancing relies on the market-making mediation of third-party digital labor platforms. However, as Blaising et al. [11] argue "...online freelancing constitutes a specific new genre of work" (p. 226:3). To this point, online freelancing is characterized by 1) knowledge-intensive work; 2) extended interaction between workers and clients; 3) project-based contracts that often have multiple deliverables; and 4) spatial and temporal flexibility [3,13,20,27,39,45,54,61].

Online freelancing tends to involve a greater reliance on conceptual skills and creativity necessary for knowledge-intensive work such as consulting and website creation [1,10,51]. Online freelancers engage in extensive communication and frequently collaborate with clients [4]. Others have also noted that the project-based nature of online freelancing tends to require that workers engage in multiple, parallel projects across different clients [11]. In addition, the lack of spatial and temporal requirements "... [allows] workers to serve multiple clients at varying hours remotely from their homes or co-working spaces instead of working full-time for a single employer" [40:1]. Online freelancing is distinct from microtasking sites such as Amazon's Mechanical Turk (AMT) by the complexity of the tasks and the expectation of worker-client interactions for guidance and sense-making [28,33,66]. As such, the combination of high skill, intensive client-worker interaction through a medium-to-long term project, and high spatiotemporal autonomy makes online freelancing a unique subset of gig work [61].

We, like other scholars, view online freelancing as a window into the future of work, given its rapid growth and projected expansion [9,40,58]. Globally, the number of registered workers on digital labor platforms tripled between 2015 and 2020, from 50 million to 163 million [40,41]. Regionally, 60 million in the U.S. participate in freelancing, up from 34% in 2015 [67,69]. The number is forecasted to reach nearly 90 million by 2028 [57]. The rapid increase in supply is paralleled by a comparable upward trend in demand. [58] highlight a 90% increase in projects offered via digital labor platforms during the period between 2016 and 2021. At a platform level, an estimated 8 million freelancers and 2.5 million businesses rely on Upwork, one of the largest freelancing platforms [70]. In sum, online freelancing is a rapidly expanding form of work within the gig economy with both clients and workers expressing a growing interest in participating in this online labor market [39,40].

Beyond the general growth of online freelancing, an active participation of women and minority workers is particularly noteworthy in digital labor. Multiple studies show that a significant percentage of women and minority workers have access to paid work through digital labor platforms [34,52]. For example, [71] highlights that women are more likely than men to identify online work as their primary source of income. High participation of workers from historically marginalized groups is also evident. According to a Pew Research Center report [68], Hispanic (30%) and Black (20%) adults work on digital labor platforms at a higher rate than White adults (12%). Despite the significant share of women and minority workers in this form of work, scholars point out that little is known about the interplay between digital labor platforms and identity positions among workers from historically marginalized groups [36,63]. As such, these workers' experiences merit further academic investigation in order to shed light on the challenges that the platform's intentional or unintentional practice poses to their livelihoods, particularly in relation to their identity management strategies, experiences, and strategies in online freelancing [7,34,36].

## 2.2 Online Freelancing: Equalizing or Reinforcing Marginalized Work

The gender wage gap in digital labor platforms is one of the most prominent manifestations of the reproduction of discrimination from the traditional workplace [7,25,32]. Indeed, [32] demonstrates that gender is a major determinant of income disparity in digital labor, with men earning considerably more per hour than women performing the same work. Likewise, [25] found that female online freelancers request just 74% of their male counterparts' hourly bill rate. Additionally, [7] revealed that women earn an average hourly wage of $28.20 per hour, while men earn an hourly wage of $45.07 on a digital labor platform.

Gendered expectations around caregiving responsibilities and household tasks also shape work patterns on digital labor platforms. [3] found that female workers' labor platform experiences are constrained by their gendered household positions. They note that childcare responsibilities, which are traditionally associated with women, lead married women with children to prefer a less cognitively demanding job with negotiable working hours, even if this means earning less income. Other researchers argue that unequal gender divisions of domestic labor influences work experiences in digital labor platforms [2]. Similarly, [36] points out women workers' commitment to childcare and domestic responsibilities often precludes them from bidding on higher-paying jobs that often require quick turnaround and/or extended periods of uninterrupted work.

Experiences of discrimination in online work are exacerbated by ostensibly neutral platform features and algorithms which in practice reflect existing societal biases [6,17,30,35,38,43,49,53,60]. For example, [17] discovered that Black English teachers frequently receive lower ratings because they receive fewer requests from new students whose parents expect White teachers on online teaching platforms. Hannák and colleagues [30] found that perceived gender and race is correlated with search ranking algorithms in a digital labor platform, potentially due to the biased client feedback toward racial and gender groups. And [43] point out that women workers are susceptible to bias, harassment, and safety incidents due to the lack of gender-sensitive platform policy and infrastructure.

Gender-based occupational segregation is also prevalent in the online labor market. Despite having similar qualifications, female workers are more likely to be hired for historically female-typed occupations (e.g., translation) than for male-typed occupations (e.g., software development) according to Galperin's findings [26]. Likewise, [64] discovered that gender roles are deeply embedded in digital labor platforms, as observed by the predominance of female workers on a home cooking chef provider platform.

In addition to the biases mediated by platforms, workers also experience discriminatory treatment from clients due to their identity attributes [16,30]. For example, [30] reveal nationality-based discrimination reflected in pay rate disparities by comparing English teachers from the U.S. and the Philippines. The study notes that perceived gender and race are significantly related to worker evaluation outcomes such as ratings and reviews. [17] notes that non-White teachers must engage in "extralinguistic labor" by answering racially sensitive questions from their students on digital labor platforms. Additionally, [6,17,30,35,38,43,49,53,60] reported disabled workers face discrimination and harassment through rating and evaluations on digital labor platforms.

In summary, in contrast to the idealist promises of digital labor as an opportunity to level discrimination and exclusion found in traditional labor markets, evidence shows that online labor extends historical marginalization and inequality. The critical views of online work enabled by platforms and algorithms presented here demonstrate the generation and reproduction of biases toward different traditionally marginalized groups, including women, minority workers, and individuals with disabilities, which have long been observed in traditional labor markets.

## 3 METHODS, DATA AND ANALYSIS

Addressing our guiding research question, this study was designed as a longitudinal panel study of online freelancers, allowing us to track changes over time [44]. We conducted semi-structured interviews with 108 US-based online freelancers over three annual rounds of data collection (December 2019 to March 2022). Our focus is on Upwork.com[2], the digital platform at the center of the online freelance labor market [21,46,54]. Upwork's online freelance market involves 8,000,000 freelancers and more than 2,500,000 businesses, with approximately 152,000 core active clients that are spending at least $5,000 on contracts per year[3]. This makes Upwork one of the largest online digital labor markets[4].

The study's longitudinal panel design provides us with an opportunity to observe the continuous evolution of the Upwork platform and analyze the experiences of freelancers who participate in this marketplace. The longitudinal design involves yearly contact with a carefully selected sample of online freelancers. It is a panel study as we speak with the same people each year. We followed the purposive sampling strategies proposed in existing literature on non-standard platform workers [55,56].

Data collection is centered on an online semi-structured interview that is complemented with a survey and secondary data collection[5]. The online interview is structured around a set of questions asked each round with other questions that may be asked once or more often. The survey is designed to support the interview, and is used to collect demographic and other quantifiable data regarding workers' freelancing experiences. The combination of repeating and topical questions allows for comparison across time while also adapting to what we are learning from each round of data collection. The interview questions, survey questions and data collection protocols are supported by secondary data collection that includes - with their permission - the freelancers' profiles on Upwork, LinkedIn and other sources, when possible. Data collection is done using a set of trained research staff.

### 3.1 Recruitment and Participants

We recruit and compensate our participants on Upwork, hiring them to complete a one-hour research job that includes both a short survey and a semi-structured interview. Participants are compensated $35 dollars for each round of data collection and are paid directly on the platform. At this point in the study, we have sustained digital contact with a total of 108 unique participants across three years of data collection (see Table 1).

Our sampling is premised on including freelancers across a diversity of personal attributes as well as different job types, skill levels, years on the platform, and levels of success. We used a purposeful sampling framework for a representative sample of participants. Demographically we sampled across gender, race, and three broad occupational groups including: 1) Administrative (e.g., data entry, transcription, virtual assistance), 2) Technology (e.g., full stack development, network security, systems administration), and 3) Creative (copywriting, graphic design, vocalist). These three broad categories encompass all the job categories on Upwork. Our participants ranged from working on the Upwork platform from 0 to 11 years with the average (mean) years on Upwork being 3.4 years. Experience with freelancing (on and off digital labor platforms) ranged from 0 to 20 years, with the average (mean) being 4.4 years.

---

[2] Upwork website: https://www.upwork.com/.

[3] See https://www.upwork.com/press/releases/53-million-americans-now-freelance-new-study-finds-2 and https://upwork.gcs-web.com/static-files/11fe6ce1-8c50-40e5-9f4e-2710de4e27c8.

[4] Other online freelancing platforms/ online labor markets include: Fiverr (https://www.fiverr.com/), Toptal (https://www.toptal.com/), Freelancer (https://www.freelancer.com/), and many others.

[5] A detailed account of the methods, including research design choices, interview and survey instruments, and study protocols are presented in (Munoz 2023).

Table 1. Participant Characteristics

| Identity Dimension | Category | Total (108 Participants) |
|---|---|---|
| Gender | Female | 63 |
|  | Male | 42 |
|  | Unknown | 3 |
| Race | Asian | 9 |
|  | Black/African American | 22 |
|  | White | 56 |
|  | Multi-racial | 13 |
|  | Other Race | 6 |
|  | Unknown | 2 |
| Ethnicity | Hispanic/Latinx | 12 |
|  | Non-Hispanic/Latinx | 93 |
|  | Unknown | 3 |
| Occupation | Administrative | 44 |
|  | Technology | 41 |
|  | Creative | 23 |
| Education | No college degree | 15 |
|  | Associates | 7 |
|  | Bachelor's | 46 |
|  | Post-graduate degree | 38 |
|  | Unknown | 2 |

Once the sampling selection was complete, the participants were randomly assigned to research project staff who reached out to the participants through the platform and completed the data collection. The participants were first asked to fill out an online survey. Once the survey was completed, the interview was scheduled through the platform and the interview was conducted online via Zoom. Interviews were recorded and the files were renamed to ensure anonymity prior to being transcribed for analysis.

### 3.2 Data Analysis

Data analysis involves three approaches: grounded theory/inductive coding, deductive coding, and thematic analysis [12,23]. We followed practical guidelines from the literature and adapted these approaches to our data needs. Given the longitudinal and large-scale qualitative design of our study, we analyzed our qualitative data in five stages, as discussed below:
1. **Inductive theme exploration:** First, three members of the research team read through the transcripts, and each completed an exploratory coding effort, informed by principles of inductive coding on a subset of transcripts [59]. Following the individual exploratory coding stage, the researchers organized and synthesized the initial meaning units into higher-level themes.

2. **Develop initial codebook:** The inductive coding effort allowed the researcher to gain familiarity and derive themes after discussions, informing the creation of an initial codebook. During this process, the researchers held weekly meetings to share reflections, discuss points of disagreement, offer alternative explanations, and further develop a set of themes.
3. **Finalize broader themes:** The themes and patterns that emerged during the preliminary coding effort were discussed among three researchers--the researchers engaged in regularly scheduled discussions to expand, merge, and alter themes in order to craft a revised codebook and ensure proper operationalization of each code. This iterative process continued until the researchers were not introducing new meaning units from the transcripts and a codebook was created to analyze the role of gender, race and other identity attributes in online freelancing.
4. **Code a subset of transcripts for coder agreement:** In the fourth stage, each of the researchers individually coded the same subset of transcripts to ensure intercoder reliability, following the finalized codebook. Each researcher individually coded 15 transcripts using NVivo[6], a common qualitative analysis tool. Then, we ran intercoder reliability measures and ensured we reached a high level of agreement across coders. The average Kappa score was: 0.89 and the average Percent Agreement was 99.83% indicating a high-level of agreement among the individual coders.
5. **Individual coding of the entire dataset:** Once these high levels of intercoder reliability were observed, each researcher was assigned a segment of the transcripts to deductively code according to the codebook developed and refined in the previous stages. This allowed us to collectively code more than 200 interview transcripts. We used NVivo to perform the individual coding for the entire dataset.

## 4 FINDINGS

Across three annual rounds of data collection from Fall 2019 through Spring 2022, a total of 78 out of the 108 participants provided responses to questions about the role of identity dimensions in their online freelancing experience. Out of these participants, 41 (53%) are female, 35 (45%) are male, and 2 did not self-disclose their gender (2%). From these respondents, 38 (49%) self-identify as White, 38 (49%) are non-White or multi-racial, and 2 (2%) did not self-disclose their race. Our data shows participants have divergent views regarding the impact of gender, race and other identity dimensions in online freelancing. For example, 56% of all participants believe that gender influences worker's experiences or outcomes in this form of work, while 46% do not (with 2% of participants showing mixed responses across different interviews).

The iterative analyses and theme development lead us to reporting six findings regarding the roles of gender, race, other identity attributes in online freelance work: 1) female freelance work is undervalued; 2) gendered occupational expectations; 3) gendered treatment; 4) shared expectations of differential values; 5) racial stereotypes and expectations; and 6) race and ethnicity as an asset. Each of the themes from our data are explained below.

---

[6] See: https://www.qsrinternational.com/nvivo-qualitative-data-analysis-software/.

## 4.1 Female Freelance Work is Undervalued

Data demonstrate that female freelance work is undervalued and underpaid on digital labor platforms (e.g., P02, P08, P13, P15, P39, P41, P45, P46, P47) [7]. This issue was raised by both male and female participants who believe women both charge and earn less on these platforms. Several female participants recognize that they ask for less money than a male counterpart would ask for. For example, one female in a technology occupation said: *"I know I have, for many years, asked for much less money than a male counterpart would or has, and I've had direct comparisons in some jobs, and it's always shocked me" (P08, Female, Tech, White)*. This participant continued to list the reasons she felt she has accepted charging and earning less for her work: "*When I looked at why I've tolerated that, what it came down to for me was I was just happy to do the work, like money wasn't the priority for me, which seems crazy, because I've always had to support myself" (P08, Female, Tech, White)*. This highlights that some women freelancers ask for and accept lower rates for their work, especially when they are satisfied with their work tasks and working arrangements. In this view, income is perceived as secondary to work satisfaction.

Beyond the trade-off of income for work satisfaction, male and female participants noted several additional reasons that females undervalue their own work. For example female workers are perceived as being more modest and are also perceived to lack the confidence to ask for or negotiate fair rates. One female in a creative occupation noted females are uncomfortable asking for a fair rate even when they know their worth. They felt females value maintaining relationships over pay, so they compromise: *"We want to find a good compromise. We want to find a situation where everyone is satisfied and I think sometimes we do undercut ourselves a bit that way…Even if my work is worth twice as much, it's hard to be that direct and say, 'You're offering me $150, but actually I'll do it for $300.' It's hard for anyone to say that, but it's something I need to get more comfortable saying and I think that is gender influenced" (P15, Female, Creative, White)*.

Gendered socialization was a recurring reason for female freelance work being undervalued. Male and female participants believe the ways females are raised and how they are expected to show up in the workplace influence how both freelancers and clients value female freelance work and contributions. For example, a male worker in a creative occupation said "*I think women are socialized, often, to value themselves less and value their contributions less. A lot of women I went to law school with are extraordinarily accomplished, but you'd never know, not just because they're modest, but it's almost as if they think that it would be impolite or inappropriate for them to assert themselves. But men do it with no problem. Men will go around and brag on themselves for hours, but a woman who is as competent, or more, is usually less likely. So I think there's this internalized feeling that 'I deserve less than whatever I might charge,' mainly because of that societal discrimination" (P47, Male, Creative, White)*. This same view of social gender differences was echoed among female participants. For example, one female compared her negotiation approach with her husband's approach: "*So, like if I'm interviewing for a project or a job, I feel like I need that and that I'm lucky to be there. So, I feel like a lot of other women are like me, we feel like we are in the position with no power. So, we don't even think. Whereas, I have watched my husband negotiate, and he knows that they want him and that he has all the power. So, of course he's going to ask for what he wants" (P67, Female, Creative, Multi-racial)*.

---

[7] We anonymized the participants and assigned them a P-number (e.g., P01, P02, etc) to include freelancers' direct quotes and incorporate their insights in this submission.

## 4.2 Gendered Occupational Expectations

A second theme emerging from our data showcases the gendered occupational expectations in online freelancing (e.g., P01, P07, P17, P19, P42, P49, P89). Workers felt there were differences in the types of work offered to, or expected of, different genders. Both male and female participants highlighted instances they believe clients were consciously and unconsciously looking for a freelancer of a given gender based on a job or occupation. One female creative worker said "*I have been completely ignored for work that I would be great at, because I have experience renovating homes. I used to work as an insurance claims adjuster, so I've literally taken courses on home construction. But I have never once been offered a writing job in that niche, and I've applied for them. And I cannot help but assume it has to do with my gender: 'What is this woman going to know about writing about home construction?'" (P19, Female, Creative, White).*

As noted by P19, this bias may be due to generalizations about the capabilities of natural skills of each gender. Clients and employers online and off have conscious and unconscious biases about the perceived abilities of each gender. These biases are more clearly pronounced in the experiences of workers in traditionally gender-stereotyped professional roles. A male administrative worker highlights this difference among lawyers "*I think that sometimes people believe that a woman may be less capable than a man, or they're perhaps deciding that they would prefer to hire a man rather than a woman and it's just more of an unconscious bias. I definitely think that may be the case with lawyers, for example, because people have this idea that they want to hire a bulldog, they want to hire somebody tough to fix their problem, and the most effective lawyers are not bulldogs, but that's what people think, and they have a hard time imagining a woman being able to do that, or they're not comfortable with women who can do that" (P49, Male, Admin, Black/African American).*

This view of gender biases was demonstrated in other occupations. For example, technical work was perceived as associated with males: "*I think there is still a bias in which people perceive men [as] more astute technically than women. There's still a bias there, particularly in business and construction fields, which I don't think is accurate at all, but I still think there's a bias there" (P89, Male, Tech, Black/African American).* Contrastingly, administrative assistant work is associated with females. One female administrative worker said "*My field is virtual assistant on Upwork, and they always tend to want a female, because females are stereotypically in those roles. I work for a law firm and in a firm, I want to say of upwards of 500 employees, we have three male secretaries, so it's rare, very rare, so I do think, at least for what I do, gender plays a role" (P17, Female, Admin, Black/African American).*

## 4.3 Gendered Treatment

A third theme in our data highlights that gender influences the overall treatment of online freelancers (e.g., P15, P19, P29, P46). Both male and female participants noted instances of clients' attitude being gender-based. For example, it appears clients were more likely to speak in a demanding or pejorative tone to female freelancers. One male creative worker noted "*Grown men will refer to a woman as a girl. They'll use that language, and it's somewhat…they'll use language that in a different way, I would characterize as pejorative, and diminutive, and it'll just show me their mindset" (P46, Male, Creative, Black/African American).*

Several female workers discussed how the power differential between clients and freelancers is even more pronounced when their clients are male. One freelancer said "*I have noticed some perhaps gender issues with some clients, where I feel like it's an intersection with culture as well. I have had some clients that I felt were quite sexist and sort of had the attitude of I'm going to order you around because I'm male. There was at least one client where I did not encourage any more work* *from that client because I just felt a little disrespected and I felt it was gender-based. 'I can demand this; you will do this for me'" (P15, Female, Creative, White).*

Another female freelancer highlighted a similar dynamic with a male client that did not value her work or accepted criticism from her as the expert. She said "*I think that the way your clients speak to you and write…like I have a client who's a smart guy, he has a master's in engineering, but he doesn't know jack about SEO and writing…But I have to be so gentle with [him]… And I can't help but think the dynamic in the relationship would be different if…like I have had to work so hard to prove myself to this guy" (P19, Female, Creative, White).*

These instances offer examples of differential treatment across genders in online freelance work. Unlike these experiences highlighted by female freelancers, there were no instances of clients' attitude or language being perceived as demanding, pejorative, or disrespectful toward male freelancers based on their gender.

**4.4  Shared Expectations of Differential Values**

A fourth theme that emerged from our data is that differential values and roles associated with genders may be influencing freelancers' work experiences (e.g., P24, P29, P31, P67). Female participants noted that differential earnings and treatment between male and female freelancers are in part due to the gendered expectations and values often associated with gender.  For example, males are more frequently associated with being the primary breadwinner and social expectations of higher income levels despite occupation, industry, or experience level. One female freelancer said "*Historically, the males are still the primary breadwinner. I don't mean that to sound archaic or anything, but I think that they have to have a certain amount. Whether it is male or female, that primary breadwinner has to have a higher rate and expectation"* said P24 (Female, Admin, White).

On the other hand, females are associated with childcare and household responsibilities. They are viewed as primary caregivers and are expected to take a backseat when it comes to their work and careers. Workers in our sample voiced their views on socially normalized expectations toward women such as prioritizing home-based responsibilities over their professional pursuits. One freelancer said "*I don't want to sound cliché, or like I'm complaining, or whatever, because I'm not, and I love being with my son and all that stuff, but I guess…and whatever…you read in the news and stuff, and typically it is the woman who would take off or be the one more with the kids, and raising them at home and stuff, so because I'm the mom and because I'm the woman of the household, I was the one that it affected my career… I think childcare is a really big deal and a heavy burden, particularly for women and for a lot of women since, right or wrong, or whatever, even today I think women are typically the ones who are just more responsible for that, or they end up being the ones that have the gap in their career, or they have to take off work" (P29, Female, Creative, White).*

However, these perceived gendered roles are not always accurate. Females often identify as primary breadwinners, dual earners, or non-care givers. Yet, male workers continue to benefit from being perceived as primary breadwinners and they continue to be more likely than females to ask for and receive fair rates. In contrast, women who are also primary breadwinners or in dual-earning households, will be less likely to ask for their worth and less likely receive fair rates based on these skewed perceptions of gender roles. One female freelancer provided a possible interpretation of this finding by noting "*I really think why women are afraid to ask the same is just how we were raised as kids, that we look up to the husband, and the dad's supposed to take care of the family. And now I have so many friends that are not married and have to take care of themselves" (P31, Female, Creative, White).*

## 4.5  Racial Stereotypes and Biased Expectations

Beyond the role of gender, there are also divergent views regarding effects of race and ethnicity in online freelancers' experiences and outcomes. Among the 35 participants who discussed experiences related to race and/or ethnicity, 24 participants (69%) felt race and ethnicity impacts workers in online freelance work, with non-White participants being nearly twice as likely to say race matters. Yet, 11 participants (31%) felt race does not impact their experience in online freelance work. Even with this split in opinion, interview data indicate race or ethnicity mediates clients' perception of workers on the freelancing platform (e.g., P46, P63, P89, P94).

Non-White participants in our sample felt that being Black/African American and/or Hispanic influences clients' perceptions. These workers noted several stereotypes that are often assigned to workers from these racial minorities including: lack of qualifications, questioning abilities for a given occupation, or making erroneous assumptions about language skills. One male freelancer noted how the client's bias toward workers' linguistic proficiency can be influenced by the workers' ethnicity. *"Well, since I'm bilingual and half Mexican, so I'm pretty sure that I expect that some people think that my English may not be good because I'm half Mexican" (P63, Male, Creative, Multi-racial, Hispanic).* This concern over language proficiency was reflected on by a female creative worker: "*When people see my last name, they go like oh, is it Latin, or whatever... I think I am the minority. I'm an editor in a male environment; I am a black woman, and I'm Latin, so people sometimes assume that my English won't be good. They assume three hundred things. I produce and write in three languages, but I'm not going to oversell myself" (P32, Female, Creative, Black/African American).*

Beyond language expectations, data highlighted other occupational stereotypes. A Black male technology worker noted how he has experienced several times that clients must adjust their perspective in order to accept his qualifications: "*My photo shows I'm African American. I think that affects things. It's just perceptions... A lot of business owners have an adjustment to seeing that the perception of the analyst isn't what they thought in their mind. It doesn't add up to them...I would say because I'm African American, a lot of people just are not ready for someone who is African American to tell them how to run their business; or they're just not ready for roles and figures of business acumen, because I do analytics and things... It takes some people a while to understand that wow, this black person really is qualified. You learn that because people disappear, and then they come back after they have maybe one or two failed efforts with somebody who [they saw as a better fit]" (P46, Male, Tech, Black/African American).*

Similarly, another freelancer highlights how clients correlate race with technical abilities. One freelancer noted that certain racial groups are more likely to be perceived as technically proficient, whilst other groups are subject to racially biased discrimination on their technical skills. He reflected on how his race might be influencing other platform experiences, including their platform metrics. *"I sometimes worry about ethnicity and the job success score, almost being like people making a correlation between those two. Like, maybe...because I feel like a lot of people have a bias that black people aren't as technically astute as folks from...like Black and Latino people aren't as technically astute as people who are, perhaps, white or East Asian, or South Asian. So I haven't had any confirmation one way or the other, but I do wonder about it sometimes" (P89, Male, Tech, Black/African American).*

Echoing client expectations related to race, freelancers described how a non-English sounding name influences clients' perceptions of workers' language skills. *"Race, I'm not sure, because always when they see my name, they're very confused, but then once they call me and they're like, 'Oh, you actually speak good English,' they seem very relieved about it, which is a little bit funny" (P94, Female, Admin, White, Hispanic).* This view highlights erroneous client assumptions based simply on a freelancer's name. Another freelancer also noted: *"As far as the name [removed], a lot of times, and there's been a few times, I don't know why this is, I've applied for a few jobs, and it will come back and say to me, 'The clients is looking for English only candidates.' I don't know why it comes back that way, because English is the only language I know" (P59, Male, Creative, White).*

## 4.2 Race and Ethnicity as an Asset

Even as some participants felt they experienced a disadvantage from their presentation of race and ethnicity on the platform, a subset of participants highlighted their racial/ethnic identity as a benefit (e.g., P06, P09, P49, P63, P82, P87). These participants felt they were able to secure certain jobs based on their identity. Participants who saw their racial/ethnic identity as a benefit noted that their identity allowed them to make a connection with the clients, or the client benefited or actively sought out to support freelancers with a specific identity. This final theme showcases the ways identity may be positively related to worker outcomes.

One female freelancer said her connection with one client was based on their shared identity as being Black: *"A couple of times I think I've gotten clients because I'm Black, and it's generally after they go with someone else, and then they come back to me… There was an older gentleman who keeps coming back to me but my prices are too high, but I think he was more comfortable because he was Black and I was Black" (P06, Female, Creative, Black/African American).* Another participant felt a connection with several clients based on a shared language and ethnicity: *"I have gotten at least two clients that hired me for multiple jobs because I spoke Spanish, so that was good. So, yeah, I would say ethnicity is a factor" (P63, Male, Creative, Multi-racial, Hispanic).*

From this perspective, race and ethnicity could be perceived as an asset in online freelance work. One freelancer said *"I think that I have definitely had several Black clients who like working with a Black lawyer, and I totally get that and respect that. I think the fact that I'm an African American man really increased my understanding with that assignment I was telling you about around race and diversity. I think that there's a sense that someone from my experience can speak really compellingly about it, as opposed to someone with less experience, or personal experience, lived experience" (P49, Male, Admin, Black/African American).*

Identity was also seen as a benefit among non-White participants in situations when a client actively sought to hire and/or support workers from racial minorities. One freelancer said *"I'm African American, and the client that I had recently now on Upwork wanted to support African American entrepreneurs… because I didn't think that that was going to work out, but when I actually had a phone conversation with her and she was telling me a little more about herself, and she was telling me what her purpose was, her agenda was, I'm like, Okay. You're being up-front. I like it" (P87, Male, Tech, Black/African American).*

Another freelancer noted the influence of minority identity on the hiring decisions of the client. A freelancer mentioned that minority workers could be given preference for employment on the platform due to clients' intentional minority hiring practices. *"Some people might be like, 'Oh, cool. I can check the box that I hired a minority' or whatever, if they will talk to me" (P09, Female, Tech, Black/African American).* This experience was viewed as the exception when the clients are consciously and intentionally making decisions based on identity criteria.

## 5 DISCUSSION

Findings make clear that the identity-based stereotypes, biases, and expectations which have long been documented in traditional offline work are visible in online freelance work. The experiences of the online freelancers in our panel study illuminate the ways in which digital labor platforms mediate identity attributes of online freelancers. These findings stand in contrast to the narrative of digital labor platforms - and the digital labor markets that they enable - as neutral actors in online freelancing.

Second, findings focus attention to the platformization of identity dimensions like gender and race in online freelancing platforms. These findings help to highlight the complexities of online mediation in freelance work. Our data show online freelancers have a mix of views about the perceived impact of these attributes in online freelancing. And, these worker's experiences showcase the different ways in which identity is reflected in online work practices.

Building from these, here we first discuss how identity biases, stereotypes and expectations are embedded into online freelance work. We then discuss how the platformization of gender, race and other attributes allowed for continued marginalization in online freelancing. We conclude the discussion with implications of our findings and provide directions for future work. We focus on these points given their centrality to contemporary HCI and CSCW scholarship [e.g, 2,11,14,19,37, 36, 52].

**5.1 Online Freelancing Reinforces Identity Stereotypes, Bias and Expectations**

While workers are divided regarding perceptions of the role and importance of gender and race within online freelance work, the lived experiences of the online freelancers in our study demonstrate the persistence of long-standing identity issues at work, including stereotyping, bias and differential expectations in online freelancing platforms. Findings highlight that both freelancers and clients perceive and assess workers differentially based on their gender, occupation, and race. We demonstrate that online freelance work embeds gender stereotypes, bias and expectations through: 1) undervaluing female freelancer's work, 2) sustaining gendered occupational expectations and preferences, 3) allowing differential treatment based on gender, and 4) enabling shared expectations of gendered values and roles. We also provide evidence of the impact of race and ethnicity, leading to worker experiences of 5) racial stereotypes and biased expectations; as well as examples of 6) race and ethnicity as an asset in online freelancing.

The findings reported here mirror existing literature on the differential experiences of workers in the gig economy based on their gender [2,24,25,26,43]. Findings from this study also demonstrate that both male and female workers tend to have similar views of gendered differences in online work and these align with previous work on gender differences at work in traditional (offline) working arrangements [31,62]. For example, the socially normalized gender expectations, such as women being humble, primary caregivers and men being assertive, primary breadwinners have been long documented in literature. Female freelancers noted how the internalization of socially normalized gender expectations affects the process and outcomes of online work such as setting and negotiating fair pay rates for their work.

We also highlight the experiences of female workers who are hesitant to assert their worth because they fear doing so would contradict socially desirable or anticipated gender roles. As a result, female workers may choose to offer their work for lower pay rates than they believe they deserve. This finding is consistent with the earlier studies that reveal female freelancers ask for significantly less per hour than their male counterparts on a digital labor platform [25]. We provide empirical explanation to this finding by revealing underlying conditions and social expectations that shape how female work is undervalued.

The consequence of socially normalized gender expectations is not limited to how much men and women earn on digital labor platforms. It also influences what jobs they could perform. Findings show workers are subjected to skill stereotyping in online freelancing. For female workers, this means that they are assumed to be more likely to possess a good quality of certain skills (e.g., communication), and thus would be better suited for a particular job than their male counterparts (e.g., virtual assistant). This gendered occupational competency affects each gender, and the types of job for which each gender is expected to apply to and be hired for. Our data shows that both male and female freelancers believe and experience clients with conscious or unconscious bias toward occupational skills based on a given gender. This finding aligns with previous research that found that female workers are more likely to be hired for traditionally female-typed occupations despite having comparable qualifications [26,64].

Findings reported here further our understanding of the roles of race and ethnicity in online freelancer's work evaluations [17,30]. We expand on how stereotypes, bias and expectations related to workers' race and ethnicity are reflected in online freelance work. By raising key concerns among non-White online freelance workers, we build on our understanding of the stereotypes and biased expectations persistent among Black and Latinx workers online. Freelance workers from these traditionally marginalized racial/ethnic identities highlighted experiences of being perceived as less competent based on their race and/or ethnicity. Workers believe clients will make assumptions about their skills and abilities, judging on stereotypes about who is "technically capable" or what type of race should be giving "business advice". The platform also allows clients to make erroneous assumptions about workers' language abilities. We provide evidence of hesitancy among clients to work with individuals who do not fit their view of someone with English proficiency, regardless of the workers actual English language proficiency.

Findings also highlight how race and gender are not mutually exclusive identity categories. Several online freelancers in our sample raise concerns and strategies to address issues of their experiences of intersectionality. One freelancer highlighted the intersecting role of their identities, including how the combination of her name, her presentation as a Black Latin woman, and her age influence how people will perceive her in freelancing platforms. "*When people see my last name, they go like oh, is it Latin, or whatever... I think I am the minority. I'm an editor in a male environment; I am a black woman, and I'm Latin, so people sometimes assume that my English won't be good. They assume three hundred things. I produce and write in three languages, but I'm not going to oversell myself*" *(P32 Female, Creative, Black/African American).* While our findings suggest intersectionality is an area of critical importance to address and minimize inequality in online freelancing, understanding the role of intersectionality within digital labor platforms and its impact on workers outcomes requires further attention.

## 5.2   Platformization of Gender, Race, and Marginalization

Findings also make clear that beyond extending traditional views of identity attributes, online freelance work leads to the 'platformization' of race, gender, and outcomes of exacerbated marginalization and inequality. Like many other scholars in CSCW, we see platformization as the process of defining workers' various identity dimensions like gender and race through technology. Indeed, prior research has highlighted how Upwork has specified a structure for how workers can perform their identity and embedded restrictions and norms for workers and clients to process these identity attributes [47]. In this way, far from being neutral actors, Upwork and other similar online freelancing platforms are designed to mediate gender, race, and other identity attributes.

 By design, all platforms embed values, norms, and expectations through their requirements, features and affordances [65]. On Upwork, identity dimensions such as gender and race are embedded in the platform through the strict guidelines and surveillance of workers' name verification, photo requirements, and language proficiency scales, among other design decisions. Each of these design decisions influence how workers are presented and perceived.

The socio-technical ecosystem of online freelancing is being influenced by these platforms so as to increase their control over the online freelance worker's experience. This means online freelancers are subjected to the platformization of their race and gender. For example, respondents noted that the requirement to provide a photo of their face allows clients to make (correct or incorrect) assumptions about identity. This challenges the view of gender and race as a non-issue. Indeed, the photo requirement can be a challenge for some workers and an advantage for others. For example, one freelancer articulated that his photos impact his success. *"…of course, coming from a position of privilege, it's really hard for me to gauge, because if you've been given that privilege your entire life, it's really hard to understand what the absence of that would look like. But I definitely think that me, being a white man, and also of course I have my picture up so they can see very clearly who I am, what I look like, I'm sure that impacts my success levels" (P41, Male, Creative, White and Asian).*

The platformization of race and gender is also manifested through the "filtering" mechanisms that are designed into the platform. For example, Upwork provides multiple features that disassemble freelancers' identity attributes into readily searchable identifiers such as location and linguistic proficiency. This filtering feature affords a context for interaction in which clients make a search and hiring decision by controlling visibility of workers based on their identities. For instance, Upwork offers a toggle button (Figure 1) that enables clients to include freelancers from the United States while excluding freelancers from other countries.

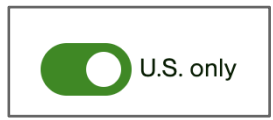

Figure 1. Upwork Offers this Option

Upwork also offers search filters for selecting languages that a worker speaks. While the platform has over 100 language options, it has a separate feature that classifies workers based only on their English proficiency (Figure 2). Our analysis shows that workers perceive English as a major attribute factored into hiring decisions. Some of our study participants encountered clients who had a specific preference for English-speaking freelancers. The platform accommodated such a preference by providing a worker search filter that categorizes workers based on their language ability. This filtering of languages can be problematic especially among workers whose identity presentation is non-conforming to the client expectations of English speakers. It is easy to imagine that this may be problematic for workers whose race/ethnicity presentation is non-White even when the English proficiency levels are high.

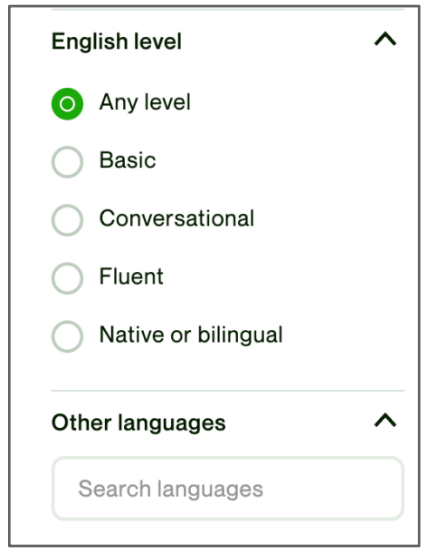

Figure 2. English Proficiency on Upwork

In addition to the platformization of gender and race, the platform also mediates marginalization. Our findings make clear that rather than addressing inequality or bias, the platform's communicative and discursive design practices drive continued marginalization. The platform's US-only filter and English proficiency level classification feature communicate preferred hiring eligibility for workers through visual signals. The prioritization of geographical location and linguistic proficiency may reflect the platform's adaptation of existing discourse, such as the economic value of English in labor markets [29]. However, such design decisions allow clients to make discriminatory hiring decisions. Simultaneously, these options create an entry barrier to workers from diverse geographical and linguistic backgrounds, potentially reinforcing the predominance of workers with pre-given or earned geographical presence and linguistic capability. Regardless of their other professional qualifications and experiences, the presence of workers whose identity presentation does not conform to such design choices is eliminated at the onset of the clients' hiring process. This calls into question the argument from some that digital labor platforms empower workers from geographically disadvantaged groups by providing them access to the global labor market [19,27,53].

### 5.3 Strategies to Take Back Control of Identity

The driving force in the platformization of race and gender is control. Without a socio-technical ecosystem that reinforces the platform's control, it risks losing its market-making position. Seen this way, the platformization of race and gender contributes to the platform's success. If online freelancers were allowed to establish a unique identity apart from the platform-controlled identity, the worker's dependency on the platform is weakened [47]. Indeed, multiple participants in the study discussed the need to over-amplify their identity by putting their photos up to avoid having negative experiences related to bias. These workers have adopted a strategy that makes their identity more visible, including on Upwork, their personal websites or other profiles. Freelancers have learned these strategies after several previous experiences of bias. Other strategies shared by participants include changing their names, or strategizing how they are being perceived on their profiles. These strategies are adopted in order for workers to align their online presentation with the expectations from clients.

Some of the online freelance workers are aware of the importance of identity presentation and seek to control aspects of their presentation that they have control over. Workers from our study actively seek to augment the visible aspects of professional persona in order to appeal to clients. While such identity presentation practices could be required for both male and female workers, women workers believe they require more effort to deploy distinct strategies to have the same effect as their male counterparts on the platform. One freelancer highlighted that *"Gender is undoubtedly one of those things that figures into that profile persona, and women who are translators may have to present themselves somewhat differently to achieve the same effect" (P47 Male, Creative, White).* This suggests an alarming possibility - that workers are facing the same established biases and marginalization from the traditional labor market but with greater constraints given the platformization of gender and race.

### 5.4 Implications and Future Work

Taken together, this study suggests that rather than neutral actors mediating work transactions, platforms like Upwork do not level the playing field among workers. Rather than reducing bias and discriminatory practices, we demonstrate that the lived experience of freelancers suggest that digital labor platforms reflect and exacerbate the gendered and racialized views of traditional work. Social identities as performed on digital labor platforms are mediating workers' experiences and their strategies for work success. Freelancers face differences across the value placed on their work, their occupational and social expectations, and their overall treatment by clients within this form of work. Findings align with critical views of online labor which suggests women and minority workers are increasingly scrutinized and marginalized, enabled by the platform design [3,16,30,43,53,60].

The issues raised in this paper provide opportunities for future research. Based on our findings, we suggest that future work explores how intersectionality [15] is mediated online and the implications for online freelancers. Black and Latinx females in our sample reflected on how their gender and race/ethnicity mediated their experiences and influenced their identity presentation strategies. Thus, work advancing intersectionality in online freelancing, and the gig economy more broadly, is important and necessary given that gender, race, and other identity attributes are interrelated, and workers with multiple burdened social identities will have unique experiences with online freelance work. Applying intersectionality within the context of digital platforms will also uncover additional implications for workers strategies for identity presentation and reveal interventions which may foster more inclusion and equality in digital labor.

There is also an opportunity for CSCW and HCI researchers and practitioners to build on this work by exploring design opportunities that would support the lives of women and minority online freelancers. Given the high engagement among these workers in the digital economy, research would benefit from centering the experiences of women and members of minority groups in design-focused research. Design research, including participatory workshops and speculative design would provide an avenue to re-imagine platforms in ways that address, rather than reflect, the challenges and disadvantages that marginalized workers have traditionally faced within work contexts. It is likely that by improving the experiences of marginalized workers in digital labor platforms, more workers will benefit and be drawn to the gig economy.

Finally, a more in depth analysis of online freelancers' experiences across the years, including the impact of COVID-19 and other market shocks would also be beneficial. Future work could explore impact and changes driven by the pandemic, and compare effects to workers from across various marginalized identities to better understand the experiences of these groups. This would provide additional insight into how challenges might have been exacerbated due to external factors beyond the control of the platform and the workers.

# 6  CONCLUSION

Amid rapid changes to work and technology, we report insights regarding how digital labor platforms are helping to change or reinforce the roles of identity attributes such as gender, race, and occupation. To do so, we draw data from a longitudinal panel study of 108 US-based online freelancers across three years of annual data collection. We contribute empirical and conceptual insight about the role of platforms in differential outcomes among online freelancers. Our findings highlight the ways platforms reinforce and exacerbate identity-based stereotypes, bias and expectations in online freelance work. We highlight the need for more attention to the role of intersectionality in online freelance work, including how multiple marginalized identities mediate workers' experiences and outcomes. We discuss the implications to the workers and future opportunities for research and design.

# REFERENCES

1.  Mats Alvesson. 2004. *Knowledge Work and Knowledge-Intensive Firms*. OUP Oxford.

2.  Ira Anjali Anwar, Joyojeet Pal, and Julie Hui. 2021. Watched, but Moving: Platformization of Beauty Work and Its Gendered Mechanisms of Control. *Proc. ACM Hum.-Comput. Interact.* 4, CSCW3: 1–20.

3.  Mohammad Amir Anwar and Mark Graham. 2020. Hidden transcripts of the gig economy: labour agency and the new art of resistance among African gig workers. *Environment & planning A* 52, 7: 1269–1291.

4.  Stephen R. Barley and Gideon Kunda. 2006. Contracting: A New Form of Professional Practice. *Academy of Management Perspectives* 20, 1: 45–66.

5.  A. Barzilay. 2019. The Technologies of Discrimination: How Platforms Cultivate Gender Inequality. *Law & Ethics of Human Rights* 13, 2: 179–202.

6.  A. Barzilay. 2019. Discrimination without discriminating: Learned gender inequality in the labor market and gig economy. *Cornell journal of law and public policy* 28, 3: 545–568.

7.  Arianne Renan Barzilay and Anat Ben-David. 2016. Platform inequality: Gender in the gig-economy. *Seton Hall law review* 47: 393.

8.  Thor Berger, Carl Benedikt Frey, Guy Levin, and Santosh Rao Danda. 2019. Uber happy? Work and well-being in the "Gig Economy." *Economic Policy* 34, 99: 429–477.

9.  Janine Berg, Marianne Furrer, Ellie Harmon, Uma Rani, and M. Six Silberman. 2018. Digital labour platforms and the future of work. *Towards Decent Work in the Online World. Rapport de l'OIT*.

10. Frank Blackler. 1995. Knowledge, Knowledge Work and Organizations: An Overview and Interpretation. *Organization Studies* 16, 6: 1021–1046.

11. Allie Blaising, Yasmine Kotturi, Chinmay Kulkarni, and Laura Dabbish. 2021. Making it Work, or Not: A Longitudinal Study of Career Trajectories Among Online Freelancers. *Proc. ACM Hum.-Comput. Interact.* 4, CSCW3: 1–29.

12. Virginia Braun and Victoria Clarke. 2006. Using thematic analysis in psychology. *Qualitative research in psychology* 3, 2: 77–101.

13. Eliane Léontine Bucher, Peter Kalum Schou, and Matthias Waldkirch. 2021. Pacifying the algorithm – Anticipatory compliance in the face of algorithmic management in the gig economy. *Organization (London, England)* 28, 1: 44–67.

14. Juan Carlos Alvarez de la Vega, Marta E. Cecchinato, and John Rooksby. 2021. "Why lose control?" A Study of Freelancers' Experiences with Gig Economy Platforms. *Proceedings of the 2021 CHI Conference on Human Factors in Computing Systems*, Association for Computing Machinery, 1–14.

15. Kimberle Crenshaw. 1991. Mapping the Margins: Intersectionality, Identity Politics, and Violence against Women of Color. *Stanford law review* 43, 6: 1241–1299.

16. Nathaniel Ming Curran. 2020. Intersectional English(es) and the Gig Economy: Teaching English Online. *International Journal of Communication Systems* 14, 0: 20.

17. Nathaniel Ming Curran. 2021. Discrimination in the gig economy: the experiences of Black online English teachers. *Language and Education*: 1–15.

18. T. R. Dillahunt, X. Wang, E. Wheeler, and H. F. Cheng. 2017. The sharing economy in computing: A systematic literature review. *Proceedings of the*.

19. Jan Drahokoupil and Maria Jepsen. 2017. The digital economy and its implications for labour. 1. The platform economy. *Transfer: European Review of Labour and Research* 23, 2: 103–107.

20. Michael Dunn, Isabel Munoz, and Mohammad Hossein Jarrahi. 2023. Dynamics of flexible work and digital platforms: Task and spatial flexibility in the platform economy. *Digital Business* 3, 1: 100052.

21. Michael Dunn, Isabel Munoz, and Steve Sawyer. 2021. Gender Differences and Lost Flexibility in Online Freelancing During the COVID-19 Pandemic. *Frontiers in sociology* 6: 738024.

22. Eurofound. 2015. *New forms of employment*. Publications Office of the European Union.

23. Jennifer Fereday and Eimear Muir-Cochrane. 2006. Demonstrating Rigor Using Thematic Analysis: A Hybrid Approach of Inductive and Deductive Coding and Theme Development. *International Journal of Qualitative Methods* 5, 1: 80–92.

24. Eureka Foong and Elizabeth Gerber. 2021. Understanding Gender Differences in Pricing Strategies in Online Labor Marketplaces. *Proceedings of the 2021 CHI Conference on Human Factors in Computing Systems*, Association for Computing Machinery, 1–16.

25. Eureka Foong, Nicholas Vincent, Brent Hecht, and Elizabeth M. Gerber. 2018. Women (Still) Ask For Less: Gender Differences in Hourly Rate in an Online Labor Marketplace. *Proc. ACM Hum.-Comput. Interact.* 2, CSCW: 1–21.

26. Hernan Galperin. 2021. "This Gig Is Not for Women": Gender Stereotyping in Online Hiring. *Social science computer review* 39, 6: 1089–1107.

27. Mark Graham, Isis Hjorth, and Vili Lehdonvirta. 2017. Digital labour and development: impacts of global digital labour platforms and the gig economy on worker livelihoods. *Transfer: European Review of Labour and Research* 23, 2: 135–162.

28. Mary L. Gray and Siddharth Suri. 2019. *Ghost Work: How to Stop Silicon Valley from Building a New Global Underclass*. Houghton Mifflin Harcourt.

29. François Grin. 2001. English as economic value: facts and fallacies. *World Englishes* 20, 1: 65–78.

30. Anikó Hannák, Claudia Wagner, David Garcia, Alan Mislove, Markus Strohmaier, and Christo Wilson. 2017. Bias in Online Freelance Marketplaces: Evidence from TaskRabbit and Fiverr. *Proceedings of the 2017 ACM Conference on Computer Supported Cooperative Work and Social Computing*, Association for Computing Machinery, 1914–1933.

31. Madeline E. Heilman. 2012. Gender stereotypes and workplace bias. *Research in Organizational Behavior* 32: 113–135.

32. Andrea M. Herrmann, Petra M. Zaal, Maryse M. H. Chappin, Brita Schemmann, and Amelie Lühmann. 2023. "We don't need no (higher) education" - How the gig economy challenges the education-income paradigm. *Technological forecasting and social change* 186: 122136.

33. Debra Howcroft and Birgitta Bergvall-Kåreborn. 2019. A typology of crowdwork platforms. *Work Employment And Society* 33, 1: 21–38.

34. Ursula Huws, Neil H. Spencer, and Simon Joyce. 2016. Crowd work in Europe feps studies. Retrieved December 27, 2022 from https://uhra.herts.ac.uk/bitstream/handle/2299/21934/crowd_work_in_europe_draft_report_last_version.pdf

35. Basileal Imana, Aleksandra Korolova, and John Heidemann. 2021. Auditing for Discrimination in Algorithms Delivering Job Ads. *Proceedings of the Web Conference 2021*, Association for Computing Machinery, 3767–3778.

36. Al James. 2022. Women in the gig economy: feminising "digital labour." *Work in the Global Economy* 2, 1: 2–26.

37. Mohammad Hossein Jarrahi, Will Sutherland, Sarah Beth Nelson, and Steve Sawyer. 2020. Platformic management, boundary resources for gig work, and worker autonomy. *Computer supported cooperative work: CSCW: an international journal* 29, 1-2: 153–189.

38. Ria Kasliwal. 2020. Gender and the gig economy: A qualitative study of gig platforms for women workers. *Observer Research Foundatio, ORF Issue Brief359*.

39. Otto Kässi and Vili Lehdonvirta. 2018. Online labour index: Measuring the online gig economy for policy and research. *Technological forecasting and social change* 137: 241–248.

40. Otto Kässi, Vili Lehdonvirta, and Fabian Stephany. 2021. How Many Online Workers are there in the World? A Data-Driven Assessment. *arXiv:2103.12648 [econ, q-fin, stat]*.

41. Siou Chew Kuek, Cecilia Paradi-Guilford, Toks Fayomi, et al. 2015. The global opportunity in online outsourcing. .

42. Ming D. Leung and Sharon Koppman. 2018. Taking a Pass: How Proportional Prejudice and Decisions Not to Hire Reproduce Gender Segregation. *American Journal of Sociology* 124, 762–813. Retrieved from http://dx.doi.org/10.1086/700677.

43. Ning F. Ma, Veronica A. Rivera, Zheng Yao, and Dongwook Yoon. 2022. "Brush it Off": How Women Workers Manage and Cope with Bias and Harassment in Gender-agnostic Gig Platforms. *Proceedings of the 2022 CHI Conference on Human Factors in Computing Systems*, Association for Computing Machinery, 1–13.

44. Isabel Munoz. 2023. Marginalization in the Future of Work: The Role of Intersectional Identities and Platforms in the Trajectories of Online Freelancers. Retrieved from https://search.proquest.com/openview/7a1f6ff17e0406c47df4c7830c16db03/1?pq-origsite=gscholar&cbl=18750&diss=y.

45. Isabel Munoz, Michael Dunn, and Steve Sawyer. 2022. Flexibility, Occupation and Gender: Insights from a Panel Study of Online Freelancers. *Information for a Better World: Shaping the Global Future*, Springer International Publishing, 311–318.

46. Isabel Munoz, Michael Dunn, and Steve Sawyer. 2022. Flexibility, Occupation and Gender: Insights from a Panel Study of Online Freelancers. *Information for a Better World: Shaping the Global Future*, Springer International Publishing, 311–318.

47. Isabel Munoz, Michael Dunn, Steve Sawyer, and Emily Michaels. 2022. Platform-mediated Markets, Online Freelance Workers and Deconstructed Identities. *Proc. ACM Hum.-Comput. Interact.* 6, CSCW2: 1–24.

48. Isabel Munoz, Steve Sawyer, and Michael Dunn. 2022. New futures of work or continued marginalization? The rise of online freelance work and digital platforms. *2022 Symposium on Human-Computer Interaction for Work*, Association for Computing Machinery, 1–7.

49. Robert A. Peterson. 2022. Heterogeneity in the US gig economy with a focus on gender. *International Journal of Applied Decision Sciences* 15, 3: 365–384.

50. Amanda Peticca-Harris, Nadia deGama, and M. N. Ravishankar. 2020. Postcapitalist precarious work and those in the "drivers" seat: Exploring the motivations and lived experiences of Uber drivers in Canada. *Organization* 27, 36–59. Retrieved from http://dx.doi.org/10.1177/1350508418757332.

51. Walter W. Powell and Kaisa Snellman. 2004. The Knowledge Economy. *Annual review of sociology* 30: 199–220.

52. Joel Ross, Lilly Irani, M. Six Silberman, Andrew Zaldivar, and Bill Tomlinson. 2010. Who are the crowdworkers? shifting demographics in mechanical turk. *CHI '10 Extended Abstracts on Human Factors in Computing Systems*, Association for Computing Machinery, 2863–2872.

53. Shruti Sannon and Dan Cosley. 2022. Toward a More Inclusive Gig Economy: Risks and Opportunities for Workers with Disabilities. *Proc. ACM Hum.-Comput. Interact.* 6, CSCW2: 1–31.

54. S. Sawyer, M. Dunn, I. Munoz, F. Stephany, and R. Raheja. 2020. Freelancing online during the COVID-19 pandemic.

55. Andrey Shevchuk and Denis Strebkov. 2018. Safeguards against opportunism in freelance contracting on the internet. *British journal of industrial relations* 56, 2: 342–369.

56. Andrey Shevchuk, Denis Strebkov, and Shannon N. Davis. 2019. The Autonomy Paradox: How Night Work Undermines Subjective Well-Being of Internet-Based Freelancers. *ILR Review* 72, 1: 75–100.

57. Ashley Stahl. 2022. Say Goodbye To The 9 To 5: How To Master The Gig Economy. *Forbes*. Retrieved November 25, 2022 from https://www.forbes.com/sites/ashleystahl/2022/09/02/say-goodbye-to-the-9-to-5-how-to-master-the-gig-economy/.

58. Fabian Stephany, Otto Kässi, Uma Rani, and Vili Lehdonvirta. 2021. Online Labour Index 2020: New ways to measure the world's remote freelancing market. *Big Data & Society* 8, 2: 20539517211043240.

59. A. Strauss and J. M. Corbin. 1997. *Grounded theory in practice*. Sage.

60. Elisabetta Stringhi. 2022. Addressing gendered affordances of the platform economy: The case of UpWork. *Internet Policy Review: Journal on Internet Regulation* 11, 1: 1–28.

61. Will Sutherland, Mohammad Hossein Jarrahi, Michael Dunn, and Sarah Beth Nelson. 2020. Work Precarity and Gig Literacies in Online Freelancing. *Work Employment And Society* 34, 3: 457–475.

62. Susan Trentham and Laurie Larwood. 1998. Gender Discrimination and the Workplace: An Examination of Rational Bias Theory. *Sex roles* 38, 1: 1–28.

63. N. Vyas. 2020. Gender inequality- now available on digital platform': an interplay between gender equality and the gig economy in the European Union. *European Labour Law Journal*.

64. Natasha A. Webster and Qian Zhang. 2020. Careers Delivered from the Kitchen? Immigrant Women Small-scale Entrepreneurs Working in the Growing Nordic Platform Economy. *NORA - Nordic Journal of Feminist and Gender Research* 28, 2: 113–125.

65. Langdon Winner. 1989. Do Artifacts Have Politics? In *The whale and the reactor a search for limits in an age of high technology*. University of Chicago Press, Chicago, 19–39.

66. Alex J. Wood, Mark Graham, Vili Lehdonvirta, and Isis Hjorth. 2019. Good Gig, Bad Gig: Autonomy and Algorithmic Control in the Global Gig Economy. *Work Employment And Society* 33, 1: 56–75.

67. 2021. *2021 Freelance Forward Economist Report*. Edelman Intelligence.

68. 2021. Racial and ethnic differences stand out in the U.S. gig workforce. *Pew Research Center*. Retrieved March 8, 2022 from https://www.pewresearch.org/fact-tank/2021/12/15/racial-and-ethnic-differences-stand-out-in-the-u-s-gig-workforce/.

69. Upwork Study Finds 60 Million Americans Freelancing in 2022. Retrieved January 14, 2023 from https://www.upwork.com/press/releases/freelance-forward-upwork-study-finds-60-million-americans-freelancing-in-2022.

70. 53 Million Americans Now Freelance, New Study Finds. Retrieved April 19, 2022 from https://www.upwork.com/press/releases/53-million-americans-now-freelance-new-study-finds-2.

71. World Employment Social Outlook 2021: The role of digital labour platforms in transforming the world of work. *ILO*. Retrieved December 29, 2022 from https://www.ilo.org/digitalguides/en-gb/story/world-employment-social-outlook-2021#introduction/women-stand.